# Cooperation in the Age of COVID-19: Evidence from Public Goods Games

Patrick Mellacher[1]

**Abstract**: Does COVID-19 change the willingness to cooperate? Four Austrian university courses in economics play a public goods game in consecutive semesters on the e-learning platform Moodle: two of them in the year before the crisis, one immediately after the beginning of the first lockdown in March 2020 and the last one in the days before the announcement of the second lockdown in October 2020. Between 67% and 76% of the students choose to cooperate, i.e. contribute to the public good, in the pre-crisis year. Immediately after the imposition of the lockdown, 71% choose to cooperate. Seven months into the crisis, however, cooperation drops to 43%. Depending on whether two types of biases resulting from the experimental design are eliminated or not, probit and logit regressions show that this drop is statistically significant at the 0.05 or the 0.1 significance level.

## 1 Introduction

During COVID-19 pandemic, numerous novel public goods game-like situations emerged in our daily lives. Optimal solutions to these situations require all citizens to choose actions that go well beyond their pure self-interest (i.e. to cooperate) and many choose voluntarily to engage in such behavior. However, the free-riding (i.e. defection) of some may have devastating effects for the community as a whole, as it frustrates efforts to contain the virus.

Most visibly, mask wearing and social distancing can be considered as public goods games (Brüne and Wilson 2020, Pejo and Biczok 2020).[2] These situations will likely accompany us for a protracted period of time. Once it is possible to do so, getting vaccinated is also a form of a public goods game (Lim and Zhang 2020). Beyond the choices of individuals, nations also face such dilemmas (Brown and Susskind 2020). Up to a certain degree, cooperation can be enforced, but most governments must rely on widespread voluntary compliance with the containment measures (Bradford et al. 2020). In addition to that, cooperative behavior is for obvious reasons also important in other contexts.

What are the effects of COVID-19 on the willingness to cooperate? Empirical evidence on other disasters is inconclusive, as natural disasters like hurricanes are on the one hand seen to trigger widespread solidarity and prosocial behavior described by Zaki (2020) as "catastrophe compassion". On the other hand, infectious diseases may have a different effect and are linked to intolerance and low levels of extraversion (Seitz et al. 2020).

I study this research question with the help of results of public goods games played by university students as part of one of their courses in public finance. I started playing this game with my students in spring 2019, which means that I have four datapoints: spring 2019, autumn 2019, at the height of the first infection wave in Austria (in the early days of the lockdown at the end of March 2020) and in the days before the second Austrian lockdown (end of October 2020). I present descriptive statistics and the results of probit and logit regressions.

---

[1] University of Graz, Graz Schumpeter Centre, patrick.mellacher@uni-graz.at
Universitätsstraße 15, 8010 Graz, Austria
[2] Defection in the action against COVID-19 may impose higher costs on other players than in a classical public goods game. Other games have thus also been used to describe the pandemic from a game theoretic perspective, e.g. a weakest-link public goods game (Caparrós and Finus 2020) and an assurance game (Taylor 2020).

In related literature, Ayers et al. (2020) find that cooperation declines during the period of beginning of March until the beginning of April using an online survey. They find that self-reported cooperative behavior decreases during this period of time. Buso et al. (2020) conduct an online experiment involving a public goods game and an ultimatum game with Italian participants in the end of April 2020. They find that cooperative behavior decreases with self-reported duration of the lockdown.

My study offers the following two contributions with respect to the aforementioned studies: a) it contains a baseline established in the year before the crisis and b) it features data from the first and the second wave of infections.

The rest of the paper is organized as follows: in the next section, I shortly summarize the public goods problem and experimental findings. I then describe how I collect the data. In section 4, I present the results. Finally, section 5 discusses the limitations of this study and concludes.

**2 Cooperation in the Public Goods Game: Theory and Evidence**

A public good is *non-rival*, i.e. everybody can benefit from it equally and *non-excludable*, i.e. nobody can be excluded from its consumption. A public goods game is a situation in which players have the option to contribute to such a good, where the individual costs of contribution exceed its individual benefits, but are at the same time lower than the sum of the benefits to all players.

A special case of a public goods game is the famous so-called "prisoner's dilemma". Economic theory predicts that fully rational agents who play a one-shot public goods game would not contribute anything to the public good, i.e. freeride, since this strategy maximizes the own outcome if the other players' strategies are given (yet unknown). The resulting Nash equilibrium is not Pareto optimal, since all agents could be better off, if all of them switched simultaneously to contribute. Figure 1 shows an illustration for the case of two players with a binary strategy space (contribute or do not contribute) where each contribution costs 3, but increases both players' payoff by 2.

| A / B | B: Contribute | B: Do not contribute |
|---|---|---|
| A: Contribute | 1/1 | -1/2 |
| A: Do not contribute | 2/-1 | 0/0 |

Figure 1: Simple 2-player public goods game

Real world results generally deviate from the theoretically expected outcome, however, for several reasons: a) agents may pursue a different target function than the maximization of one's own payoff, e.g. because they are kind, altruistic or are embarrassed to choose the "selfish" option, b) agents may not be fully rational, i.e. do not understand that it is a dominant strategy to do not contribute, or c) because they might be able to form more or less binding contracts if they i) are able to communicate before or during the game, and ii) are able to observe the strategies chosen by other players in some way during the game or if they are able to punish other players following an ex-post revelation of each player's contribution.

In a laboratory setting, c) is typically excluded (if it is not to be studied explicitly, see e.g. Oprea et al. 2014), even though contracts and punishment for non-cooperation seem to be highly important in the real world, think of e.g. organized crime solving the prisoner's dilemma in the real-world.

Marwell and Ames (1981) show with a series of public goods experiments that the free-riding hypothesis generally does not hold, even though it fits much better to graduate students of economics than to the general public. Andreoni (1995) finds that while on average 75% of the participants in his experiment choose to cooperate, about half of this level of cooperation can be attributed to each a) kindness and b) "confusion" about the payoff structure. Fischbacher et al. (2001) find that about 30% of the participants of their study can be classified as pure free-riders, whereas the others are willing to cooperate to a certain extent and especially under certain conditions, namely that the others contribute as well (i.e. are "conditional cooperators").

**3 Method and data**

I use data from Austrian university students who participate in an introductory course on public finance. In the third lecture, they are introduced to the public goods game and its characteristics (including the Nash equilibrium). At the end of that lecture, they are encouraged to play a public goods game via the e-learning platform Moodle for which they can gain a small number of points that count towards their final grade. In 2020, the third lecture was taught via online videos. Nevertheless, the homework was introduced after sending a link to the lecture (March) or reminding students verbally to watch the respective video (October). In each semester, I additionally sent out a reminder via e-mail to all students.

When playing the game, students are presented with the following question, where x is 0.1 in the spring 2019 and 0.2 in the other courses – accounting for a lower number of participants in these courses.

> "You are offered one participation point. You can choose, whether you want to keep it for yourself or invest it. If you invest your point, all students who answer this question (including you, as well as those, who do not invest) earn x points."

Students are allowed to change their choice and the cutoff time is set at midnight in order to avoid decisions which are enforced by students playing the game openly together. Data is available for four terms ranging from spring 2019 until autumn 2020. Table 1 shows the respective timeframes and the number of students who played the public goods game in each period (players) as well as the number of students who took the mid-term exam (active participants).

**Table 1: Players**

| Month | Players / active participants |
|---|---|
| March 2019 (21.03.-03.04.) | 42 / 45 |
| October 2019 (15.10.-21.10.) | 17 / 19 |
| March 2020 (18.03.-31.03.) | 28 / 28 |
| October 2020 (20.10.-31.10.) | 14 / 16 |

The two games played in 2020 fell accidentally into highly interesting periods of time, as the first Austrian lockdown came into effect on the 16$^{th}$ of March 2020 and the "light lockdown" on the 3$^{rd}$ of November 2020.

I identified two detectable sources of possible bias from the experimental setup:

1.) If students chose "not to invest", they got a feedback from Moodle that their answer was not correct (part 2 of the following quote), although I tried to mitigate the effect of this negative feedback by introducing a custom caption (part 1 of the following quote) explaining that their answer was in fact not wrong.

> "You chose to invest your point! That is not a wrong answer. You will gain points based on how many people chose to invest their points, which will be revealed in the next lecture
>
> The correct answer is: do not invest"

Nevertheless, two students of one semester (March 2020) changed their answer directly after receiving this feedback, which indicates that it did have an effect on them.

2.) A small number of students had to repeat the course and thus played the game twice. The average contribution in the second game is lower than that of the first (though larger than 0), aligning well with the literature on falling contribution for repeated public goods games. Fischbacher and Gächter (2010) provide evidence that this behavior can be traced back to the fact that most people prefer to contribute less than others and the own contributions therefore directly depend on their belief about what other players will contribute. One could thus hypothesize that the fact that some players switched from contribution to non-contribution in their second game can be traced back to their belief that other players will contribute as much as their colleagues in the previous semester, and that the expected level of cooperation is too low to justify cooperation.

When I account for these biases, I remove the respective players (e.g. all players who played the game for the second time) from the dataset. In order to protect students' privacy, I do not provide descriptive statistics on bias type 2, but accounting for the bias only has little effect on the regression results, as shown in the next section.

**4 Results**

Figure 2 reports the average cooperation of the public goods game. Between 67 and 76% of the students choose to cooperate, i.e. contribute to the public good, in the pre-crisis year. Immediately after the imposition of the lockdown, 71% choose to cooperate. Seven months into the crisis, however, cooperation drops to 43%. After accounting for the bias of type 1 as described in the previous section, this negative effect is even more pronounced. Cooperation in March 2020 is then at 77%, i.e. slightly, but not statistically significantly higher than in the experiments conducted in 2019.

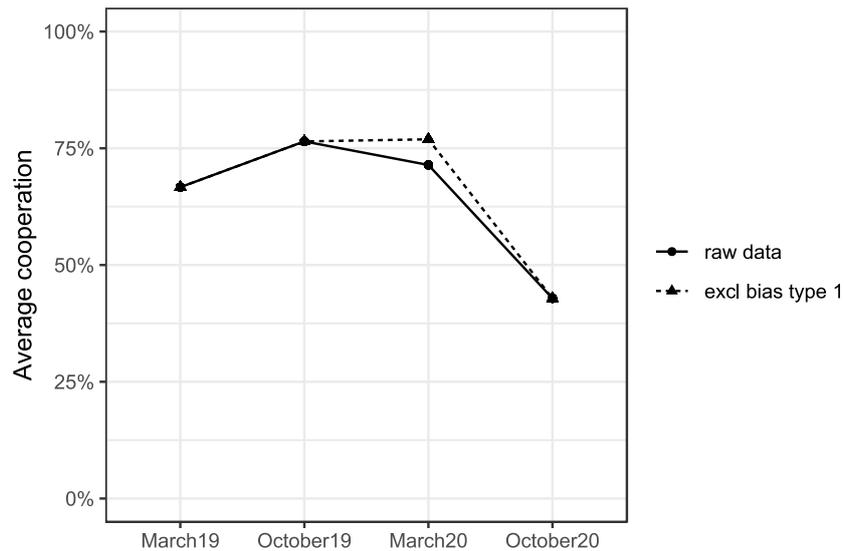

**Figure 2**: Cooperation over time

Logit (table 2) and probit (table 3) regressions finally show that this fall is statistically significant at the 10% level for the raw data (regression 1) and at the 5% level, if one accounts for bias 1 (regression 2) or 1+2 (regression 3).

Table 2: Results of logit regressions

|  | (1) | (2) | (3) |
|---|---|---|---|
| October 2020 | -1.140* | -1.221** | -1.247** |
|  | p = 0.053 | p = 0.040 | p = 0.036 |
| Constant | 0.853*** | 0.933*** | 0.959*** |
|  | p = 0.0003 | p = 0.0002 | p = 0.0001 |
| Note: | | *p<0.1; **p<0.05; ***p<0.01 | |

Table 3: Results of probit regressions

|  | (1) | (2) | (3) |
|---|---|---|---|
| October 2020 | -0.708* | -0.756** | -0.771** |
|  | p = 0.053 | p = 0.040 | p = 0.036 |
| Constant | 0.528*** | 0.576*** | 0.591*** |
|  | p = 0.0002 | p = 0.0001 | p = 0.0001 |
| Note: | | *p<0.1; **p<0.05; ***p<0.01 | |

## 5 Discussion and conclusion

I explored whether and how the prolonged COVID-19 crisis changes the willingness to cooperate by analyzing the results of public goods games played by university students in an Austrian economics course. By comparing results from March 2020 (during the first Austrian lockdown) and October 2020 (in the days before the second Austrian lockdown) with the results of the pre-crisis year, I do not find a statistically significant effect during the first lockdown. However, I do find a strong and statistically significant fall in the willingness to cooperate in October 2020.

Four hypotheses to explain these results arise from previous research on contributions in a public goods game e.g. by Andreoni (1995), Fischbacher et al. (2001) and Fischbacher and Gächter (2010):

H1: The level of "confusion" (to use Andreoni's 1995 terms) was lower than in the previous courses, which means that the contributions traced back to this cause were lower, whereas contributions due to "kindness" remained unchanged. This case seems to be unlikely, as the theory on public goods games was discussed in class right before playing in the two games played in the pre-crisis year, which should have served to reduce the level of confusion.[3] In March 2020, the public goods game was explained with an online video accessible from the start of the game. This change in procedure was necessary due to the COVID-19 regulations, but seemingly did not to have any impact on the results. In October 2020, students were also able to watch the videos on public goods games before actually playing it.

H2: The crisis did not change the players' preferences, but their beliefs in a way that cannot be attributed to COVID-19. More specifically, it could be argued that students had less chances to meet each other and build trust in their fellow students during the crisis than before. This case, however, also seems to be unlikely, as a) this is a course taught in a foreign language (English) and the participants thus consisted to a large degree of exchange students who were not familiar with each other before or beside the course, and b) in both courses taught in 2020, students met exactly twice in person before switching to distance learning, i.e. they had an equal chance to get to know their classmates before the game in March and October 2020 with very different results.

H3: The crisis had a fundamental psychological impact that changed players' preferences. In the terms used in previous experimental studies, people who were altruists or conditional cooperators before COVID-19 became free-riders. Ayers et al. (2020) show that the agreement to the statement that "helping someone in need is the right thing to do" declined from the beginning of March until the beginning of April for the participants of their online survey, which may point to this hypothesis.

H4: The crisis did not change players' preferences, but their beliefs in a way that can be attributed to the COVID-19 crisis. The Austrian government framed the battle against COVID-19 in summer until early autumn 2020 primarily as a matter of individual responsibility. However, by the end of October it had become apparent that these appeals had failed and a second lockdown was imminent. Since most people are shown to be "conditional cooperators" in public goods lab experiments by e.g. Fischbacher et al. (2001), players may have extrapolated their beliefs about the average contribution from their experience with the COVID-19 crisis (where a lot of people did and do not behave cooperatively on their own).

This study is limited by the fact that it was not conducted in a laboratory setting and cannot attribute the fall in the willingness to cooperate to a particular cause. Nevertheless, its finding is significant as cooperation seems to be crucial in the combat against COVID-19 as citizens have to face many public goods game-like situations in the everyday life of the pandemic.

Further research is necessary to confirm whether this effect also holds for laboratory settings, across social and national boundaries and whether it persists in the long term. If this effect can be replicated in further studies, hypotheses 1 and 2 can likely be rejected. Using targeted surveys and comparisons with previous laboratory studies, subsequent research can also investigate whether hypotheses 3 and/or 4 can be supported.

---

[3] From the feedback received after presenting the results, however, it was apparent that "confusion" about the payoff structure did have at least a small impact.

**Declaration of Competing Interest**

None.

**Acknowledgements**: I thank my students for participating actively and thinking critically in my classes.